\documentclass[twocolumn,pre,superscriptaddress,aps,showpacs,nofootinbib]{revtex4}

\usepackage[dvips]{graphicx}
\usepackage{subfigure}
\usepackage{epsfig}
\usepackage{amsmath}

\def\be{\begin{equation}}
\def\ee{\end{equation}}
\def\bea{\begin{eqnarray}}
\def\eea{\end{eqnarray}}

\begin{document}
\title{Comparative merits of the memory function and dynamic local field correction \\ of the classical one-component plasma}
\author{James P. \surname{Mithen} }\email{james.mithen@physics.ox.ac.uk}
\affiliation{Department of Physics, Clarendon Laboratory, University of Oxford, Parks Road, Oxford OX1 3PU, UK}
\author{ J\'er\^ome \surname{Daligault}}
\affiliation{Theoretical Division, Los Alamos National Laboratory, Los Alamos, NM 87545}
\author{Gianluca Gregori}
\affiliation{Department of Physics, Clarendon Laboratory, University of Oxford, Parks Road, Oxford OX1 3PU, UK}

\date{\today}

\begin{abstract}
The complementarity of the liquid and plasma descriptions of the classical one-component plasma (OCP) is explored by studying wavevector and frequency dependent dynamical quantities: the dynamical structure factor (DSF), and the dynamic local field correction (LFC).
Accurate Molecular Dynamics (MD) simulations are used to validate/test models of the DSF and LFC.  Our simulations, which span the entire fluid regime ($\Gamma = 0.1 - 175$), show that the DSF is very well represented by a simple and well known memory function model of generalized hydrodynamics.
On the other hand, the LFC, which we have computed using MD for the first time, is not well described by existing models.
\end{abstract}


\maketitle
\section{Introduction}
The classical one-component plasma (OCP) is a standard model in the study of strongly coupled plasmas,   
playing a conceptual role similar to that of the hard-sphere model in the theory of simple liquids.
It is often used as a model of matter under extreme conditions, e.g. compact astrophysical objects.
The OCP consists of a system of identical
point charges $Ze$ with mass $m$, interacting through the Coulomb potential, and immersed
in a uniform background of opposite charge.  
In equilibrium, the system is characterised by the dimensionless coupling parameter $\Gamma=(Ze)^{2}/ak_{B}T$,
where  $a=\left(4\pi\/n/3\right)^{-1/3}$ is the mean interparticle distance
with $n$ the particle density and $T$ the temperature.

As $\Gamma$ increases, the OCP changes from a nearly collisionless, gaseous regime for
$\Gamma \ll 1$ through an increasingly correlated, dense fluid regime in which the system shares certain properties 
with ordinary liquids.
In particular, for $\Gamma > 50$ it has been found that the transport coefficients (diffusion, viscosity) of the OCP obey universal laws satisfied by dense ordinary liquids \cite{Daligault}.  
Other features of the OCP dynamics are not shared by ordinary liquids.
Most notably, because of the long range Coulomb interactions, the system exhibits the characteristic behavior of plasmas: density imbalances lead to high frequency plasma oscillations, rather than low frequency sound waves.  These high frequency plasma oscillations, not encountered in ordinary liquids, led Baus and Hansen to question the validity of the hydrodynamic limit of the OCP \cite{BausHansen}.
In fact, it was recently shown that the hydrodynamic limit of the OCP is not applicable, even at large $\Gamma$ values where high collisionality due to caging leads to liquidlike properties \cite{Mithen}.
It is the fact that the OCP shares some, but not all, properties with ordinary liquids that makes it a challenging yet fascinating system to study.

In this paper we will explore the complementarity of the liquid and plasma descriptions of the OCP by studying the wavevector and frequency dependent dynamical structure factor (DSF), $S(k,\omega)$.
The DSF contains complete information of the system dynamics at and near thermal equilibrium and is an important quantity because of its connection to inelastic light and neutron scattering experiments \cite{HansenMcdonald,BalucaniZoppi}.
Two main approaches have been proposed for modeling the DSF in the fluid regime $\Gamma < 175$: the memory function approach and the dynamic local field correction (LFC) approach.  
Largely due to the lack of `exact' results (from numerical simulations) to compare to theoretical models of the memory function and LFC, it is not clear which of these approaches is more suitable for providing a description that is simple and effective for a wide range of conditions.  
The purpose of this paper is to clarify this problem.

The memory function approach - widely used for normal liquids - represents a generalized hydrodynamics in which both equilibrium properties and transport coefficients that appear in the conventional hydrodynamic (Navier-Stokes) description are replaced by suitably defined wavevector and frequency dependent quantities.  In this approach, the DSF is written in the form \cite{BalucaniZoppi}
\begin{equation}
\frac{S(k,\omega)}{S(k)} = \frac{1}{\pi}\frac{<\omega_k^2>k^2\phi^{'}(k,\omega)}{[\omega^2 - <\omega_k^2> - \omega k^2 \phi^{''}(k,\omega)]^2 + [\omega k^2\phi^{'}(k,\omega)]^2}\,,
\label{skwequation}
\end{equation}
where $S(k)$ is the static structure factor and $<\omega_k^2> = \frac{k_B T}{m}\frac{k^2}{S(k)}$.  
The quantities $\phi^{'}(k,\omega)$ and $\phi^{''}(k,\omega)$ are respectively the real and imaginary parts of the Laplace transform of the memory function $\phi(k,t)$. 
In order to fully specify the DSF a model for the memory function is required.  
Of particular note is the Gaussian memory function model first applied to the OCP by Hansen et al. \cite{Hansen}, which looked promising at the time of their study.

The dynamics of the OCP can instead can be described in terms of the so-called dynamic local field correction (LFC), $G(k,\omega)$. This approach is more common to Coulomb systems e.g. the quantum electron gas \cite{Kugler}.
The LFC is defined by its relation to the density response function of the system, $\chi(k,\omega)$ \cite{Ichimarurev,Ichimarubook},
\begin{equation}
\chi(k,\omega) = \frac{\chi_0(k,\omega)}{1 - v(k)[1 - G(k,\omega)]\chi_0(k,\omega)}\,.
\label{lfcdef}
\end{equation}
Here $v(k) = 4\pi (Ze)^2 / k^2$ is the Fourier transform of the Coulomb potential and $\chi_0(k,\omega)$ is the density response function of an ideal gas, defined in Sec. \ref{dynamiclfc}. 
While the memory function is designed to extend the conventional hydrodynamic equations to finite wavevectors, the LFC is designed to correct the deficiencies of the mean field approximation (i.e. the Vlasov equation for the single particle distribution function, which describes the plasma oscillations but neglects any non-ideal or `collisional' effects).
That is, setting $G(k,\omega) = 0$ gives the mean field approximation for the density response function; this gives a good description of the OCP dynamics in the weak coupling regime, $\Gamma \ll 1$, only.
A non-zero $G(k,\omega)$ represents correlation effects beyond the mean field approximation.
Models for the LFC have been proposed by Tanaka and Ichimaru \cite{Ichimaru} and by Hong and Kim \cite{Hong}, but these models have barely been tested other than for a very few conditions in the original studies (and even for these conditions it was not clear how well the models agreed with the MD data).

Since the density response function and DSF are related through the fluctuation-dissipation theorem,
\begin{equation}
S(k,\omega) = -\frac{k_BT}{\pi n \omega}\Im m\{\chi(k,\omega)\}\,,
\label{flucdiss}
\end{equation}
the LFC $G(k,\omega)$ is clearly related to the memory function $\phi(k,\omega)$, albeit in a non-trivial way.
In this paper we show that the memory function is a simpler quantity to model than the LFC.  That is, a basic model for the memory function can describe both mean field and collisional effects that are characteristic of the DSF of the OCP, wheras an LFC that achieves this is much more complicated.
Specifically, as shown in Sec. \ref{memoryfunction}, the Gaussian memory function model initially proposed by Hansen et al. reproduces the MD data for the DSF to remarkable accuracy across the entire fluid regime, and for all wavevectors $k$.  In fact, the properties of the OCP mean that the model works even better than would be expected in the case of normal fluids.
On the other hand, as shown in Sec. \ref{dynamiclfc}, the LFC has a more complex structure - for this reason it is
 not well described by the models mentioned previously.
In order to reach these conclusions, we have performed highly accurate, large scale, state of the art molecular dynamics (MD) simulations for the intermediate scattering function $F(k,t)$, and from this the dynamical structure factor, $S(k,\omega)$ \cite{HansenMcdonald}, for a large number of $\Gamma$ values spanning the entire fluid regime ($0.1$,$0.3$,$1$,$5$,$8$,$9$,$9.5$,$10$,$11$,$50$,$120$,$160$,$175$). 
We have used this new data to compute the LFC of the OCP with MD for the first time: 
calculation of $G(k,\omega)$ requires very accurate MD data which was not available before now.
To conclude our study of OCP dynamics (Sec. \ref{negdispersion}), we have extracted from our MD data the value of $\Gamma$ at which `negative dispersion' of the OCP plasmon mode sets in; very recently there has been renewed interest in this particular aspect of OCP dynamics \cite{Arkhipov}.

\section{Memory Function Model}
\label{memoryfunction}
The memory function expression in Eq. (\ref{skwequation}), which can be used to represent the DSF of any single component fluid, can be shown to be an exact result \cite{BalucaniZoppi}.  In the case of the OCP, the ubiquitous plasmon peak in the DSF is ensured by the long wavelength (small $k$) behaviour of the $<\omega_k^2>$ term in the denominator of Eq. (\ref{skwequation}); as $k \rightarrow 0$, $S(k) \rightarrow k^2/k_D^2$ \cite{BausHansen}, where $k_D^2 = 3\Gamma / a^2$, and hence $<\omega_k^2> \rightarrow \omega_p$.  This small $k$ behaviour of $S(k)$ is an essential distinction between OCP statics and those of an ordinary fluid - in the latter case, $S(k)$ approaches the isothermal compressibility of the fluid in the limit $k \rightarrow 0$, which gives rise to a sound wave (rather than a plasma wave) at long wavelengths \cite{BausHansen}. 

The memory function model first applied to the OCP by Hansen et al. \cite{Hansen} consists of using the following Gaussian ansatz for the memory function,
\begin{align}
k^2\phi(k,t) &= k^2\phi(k,0)\exp(-\pi t^2/4\tau_k) \nonumber \\
             &= [\omega_L^2(k) - <\omega_k^2>]\exp(-\pi t^2/4\tau_k)\,,
\label{Gaussianansatz}
\end{align}
where the initial value of the memory function is known exactly \cite{BalucaniZoppi} and  $\omega_L^2(k) = <\omega^4>/<\omega^2>$ is given in terms of the frequency moments of $S(k,\omega)$
\begin{equation}
<\omega^n> = \int_{-\infty}^{\infty}\omega^nS(k,\omega)d\omega\,.
\end{equation}
Expressions for $<\omega^0>$, $<\omega^2>$ and $<\omega^4>$ in terms of the static structure factor $S(k)$ and the radial distribution function $g(r)$ for the OCP are given in the Appendix. Here $\tau_k$, appearing in Eq. (\ref{Gaussianansatz}), is a wavevector dependent relaxation time.
According to Eq. (\ref{Gaussianansatz}), the real and imaginary parts of the Laplace transform of the memory function are given by, respectively \cite{Ailawadi,Hansen},
\begin{equation}
k^2\phi^{'}(k,\omega) = [\omega_L^2(k) - <\omega_k^2>]\tau_ke^{-\tau_k^2\omega^2/\pi}
\label{realphi}
\end{equation}
and
\begin{equation}
k^2\phi^{''}(k,\omega) = \frac{2\tau_k}{\sqrt{\pi}}[\omega_L^2(k) - <\omega_k^2>]D(\tau_k\omega/\sqrt{\pi})\,,
\label{imagphi}
\end{equation}
where the Dawson function $D(x) = \exp(-x^2)\int_0^x\exp(y^2)dy$
\cite{Dawson}.  


\subsection{Comparison between model and MD data}
\label{skwcomp}

The parameters $<\omega_k^2>$ and $\omega_L^2(k)$ that appear in the model can be obtained by computing $S(k)$ (or equivalently $g(r)$) with MD and using the formulae given in the Appendix for the frequency moments.  The model then reduces to the determination of a single $k$ dependent parameter $\tau_k$.  The approach taken by Hansen et al. was to treat $\tau_k$ as a parameter to be fitted to the MD spectrum 
of $S(k,\omega)$.

\begin{figure}[h!]
\includegraphics{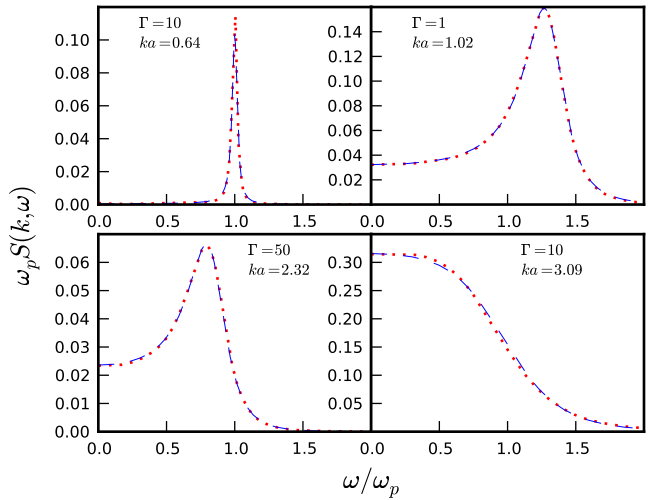}
\caption{(color online) MD results at selected $\Gamma$ and $ka$ values (dots) and the result of the Gaussian model when the parameter $\tau_k$ is fitted to the MD spectrum (dashed line).}
\label{1param}
\end{figure}

\begin{figure}[h!]
\includegraphics{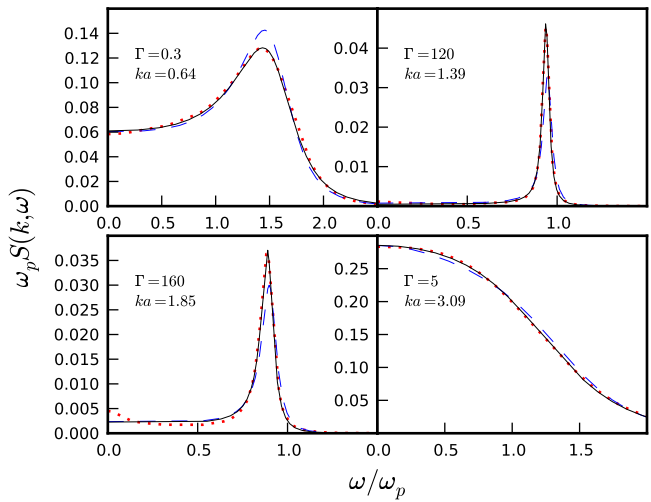}
\caption{(color online) MD results at selected $\Gamma$ and $ka$ values (dots) along with the result of the Gaussian model when only the parameter $\tau_k$ is fitted to the MD spectrum (dashed line), and the result when $<\omega_k^2>$, $<\omega_L^2(k)>$ and $\tau_k$ are all fitted (solid line).}
\label{1param2}
\end{figure}

When we do this, we find in general that the model matches the MD data remarkably well for all $\Gamma$ and $k$ values (see Fig. \ref{1param}).  In some cases, however, there are small discrepancies between the model and MD data, despite the fact that the model recreates the shape of the MD data very well (see Fig. \ref{1param2}).  Therefore, in order to determine whether these discrepancies are due to deficiencies in the model or inaccuracies in the parameters  $<\omega_k^2>$ and $<\omega_L^2(k)>$ when computed with MD, we have  separately fitted the model to the MD spectrum using all three parameters $<\omega_k^2>$, $<\omega_L^2(k)>$ and $\tau_k$.  As shown in Fig. \ref{1param2}, this three parameter fit is an even better match to the MD data.
Since the values of the parameters $<\omega_k^2>$ and $<\omega_L^2(k)>$ from the three parameter fit agree very closely (within $10\%$) with those computed with MD, we conclude that the improvement in the agreement between the model when all three parameters are fitted versus when only one is fitted is due to small inaccuracies when $<\omega_k^2>$ and $\omega_L^2(k)$ are taken from the MD $g(r)$ and $S(k)$; the model is rather sensitive to the precise values of the frequency moments.
That is, the one parameter fits are irrelevant as their comparison with the MD data is not indicative of the quality of the model.
In Figs. \ref{fig1} and \ref{fig2}, we show only the model results for when all three of the moments are used as fitting parameters.

As shown in Figs. \ref{fig1} and \ref{fig2}, at sufficiently small $k$, the MD data for $S(k,\omega)$ exhibits a sharp plasmon peak at $\omega \approx \omega_p$ for all coupling strengths $\Gamma$.  
As $k$ increases, the plasmon peak broadens until, at high $k$, $S(k,\omega)$ reduces to a single central peak at $\omega = 0$.  The model accounts remarkably well for this entire evolution, particularly for $\Gamma \leq 50$ (Fig \ref{fig1}).  At higher values of $\Gamma$, the MD data does show some additional structure at intermediate $k$ values ($ka = 2.32 \mbox{ and } 3.09$) that the model cannot reproduce.
For $\Gamma \geq 120$, a two peak structure
is visible for $ka = 2.32$ and a three peak structure for $ka = 3.09$ (Fig. \ref{fig2}).  
The small high frequency peak for $ka = 3.09$ is of particular interest - it does not appear to have been 
seen or commented upon in previous MD calculations.  
We believe that this peak is due to microscopic `caging' effects.  That is, at these lengthscales, the relatively high frequency oscillations of individual particles in the cages produced by their neighbors are imprinted on $S(k,\omega)$.
This deduction is supported by previous work showing that for $\Gamma \geq 100$, a high frequency peak appears in the velocity autocorrelation function at $\simeq 0.9\omega_p$ \cite{Daligault}; this is exactly the position of the additional peak in $S(k,\omega)$.
We note that although the model does not fully capture the additional structure in 
the MD data for these conditions, on average it does give a good account of the overall shape of the spectrum.

\begin{figure}[h!]
\includegraphics{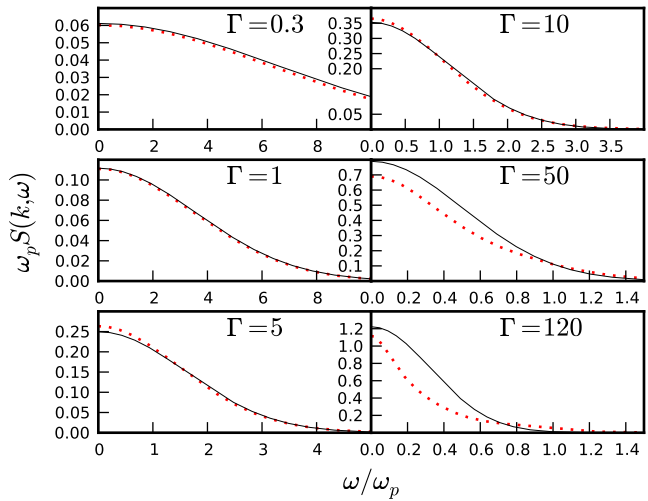}
\caption{(color online) MD data for $ka = 6.19$ (dots) and the ideal gas limit given by Eq. (\ref{idealgas}) (solid line).}
\label{idealfig}
\end{figure}

\begin{figure}[h!]
 \includegraphics{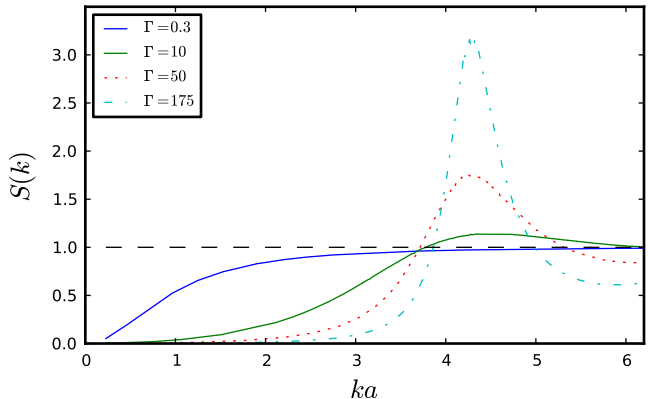}
 \caption{(color online) The static structure factor $S(k)$ for a
   range of $\Gamma$ values.  Clearly for $\Gamma > 10$ $S(k)$ is
   still oscillating at our highest $ka$ value.
 }
 \label{static}
 \end{figure}

As $k$ increases further, $S(k,\omega)$  begins to reduce to its ideal gas limit $S^{0}(k,\omega)$ \cite{Hansen,HansenMcdonald}, given by
\begin{equation}
S^{0}(k,\omega) = \left(\frac{m}{2\pi k_B T k^2}\right)^{1/2}\exp\left(-\frac{m\omega^2}{2k_BTk^2}\right)\,.
\label{idealgas}
\end{equation}
As shown in Fig. \ref{idealfig},  for $\Gamma \leq 10$, $S(k,\omega)$ is already close to $S^{0}(k,\omega)$ for our highest $k$ value ($ka = 6.19$).  For $\Gamma > 10$, significant differences appear - these differences become greater as the coupling strength increases.
This is to be expected, as at these higher coupling strengths oscillations the static structure factor $S(k)$ persist well beyond $ka = 6.19$ (Fig. \ref{static}).

In any case, as shown in Figs. \ref{fig1} and \ref{fig2}, the Gaussian model works well  at our highest $k$ value of $ka = 6.19$, regardless of whether or not this $k$ value is sufficiently large for $S(k,\omega)$ to be close to its ideal gas limit.

\section{Dynamic Local field Correction}
\label{dynamiclfc}
In the dynamic local field correction (LFC) approach, as given in Eq. (\ref{lfcdef}), the dynamics are described with reference to the density response function of an ideal gas,
\begin{equation}
\chi^{(0)}(k,\omega) = -\frac{n}{k_BT}Z\left(\sqrt{ \frac{m}{2 k_BT} } \frac{\omega}{k}\right)\,,
\end{equation}
where $Z(x) = (1 - 2xD(x)) + i\sqrt{\pi}x \exp(-x^2)$ and $D(x)$ is the Dawson function introduced in Sec. \ref{memoryfunction}.

As mentioned previously, setting $G(k,\omega) = 0$ gives the mean field approximation for the density response function; this only gives a good description of the OCP dynamics in the weak coupling regime $\Gamma \ll 1$.
A non-zero $G(k,\omega)$ represents correlation effects beyond the mean field approximation.
One commonly used approximation is to replace the LFC by its $\omega = 0$ value $G(k,0) \equiv G(k)$.  The static local field correction $G(k)$ is related to the static structure factor $S(k)$ by
\begin{equation}
G(k) = 1 + \left[1- \frac{1}{S(k)}\right]\frac{n }{k_BT}\frac{1}{v(k)}\,.
\label{gkeq}
\end{equation}
An alternative scheme is to replace the dynamic local field correction by its high frequency limit 
\begin{equation}
G(k,\infty) \equiv G(k,\omega \rightarrow \infty) = 2I(k)\,,
\label{gkwinfeq}
\end{equation}
where $I(k)$, which depends on the radial distribution function $g(r)$ of the OCP, is given in the Appendix.
Replacing $G(k,\omega)$ by either $G(k)$ or $G(k, \infty)$ results in a mean field approximation with an effective potential; it is well known that this type of scheme gives only a marginal improvement over the conventional mean field approximation \cite{Hansen}.
Thus, in order to describe well the dynamics of the OCP for $\Gamma \geq 1$, it is necessary to take collisions into account by having a frequency dependent dynamic local field correction.

For the classical one-component plasma, two main formulations of the frequency dependent local field correction have been given.
The expression given by Tanaka and Ichimaru \cite{Ichimaru}, based on their viscoelastic formalism, interpolates between the known zero frequency and high frequency limits given in Eqs. (\ref{gkeq}) and (\ref{gkwinfeq}),
\begin{equation}
G(k,\omega) = \frac{G(k) - i \omega \tau_M(k)G(k,\infty)}{1 - i\omega\tau_M(k)}\,.
\label{ichimarueq}
\end{equation}
In their prescription for computing the relaxation time $\tau_M(k)$, Tanaka and Ichimaru considered either a Gaussian or Lorentzian
ansatz \cite{Ichimaru}.  In both of these cases, they used a kinetic equation to relate the shear viscosity to the local field correction; the unknown parameter appearing in $\tau_M(k)$ was then chosen such that the estimates of the shear viscosity available from MD at the time were matched as closely as possible (see \cite{Ichimaru} for further details).

The other formulation, given by Hong and Kim \cite{Hong}, generates successive approximations for the LFC.  The first order approximation is simply to replace $G(k,\omega)$ by $G(k)$.  The second order approximation is
\begin{equation}
G(k,\omega) = G(k) - \frac{1}{2}[G(k) - G(k,\infty)]Q\left(\sqrt{ \frac{m}{2 k_BT} } \frac{\omega}{k}\right)\,,
\label{hongkim}
\end{equation}
where $Q(x) = 1/Z(x) + 2x^2 - 1$.  Because $Q(0) = 0$ and $Q(x \rightarrow \infty) = 2$, like the model of Tanaka and Ichimaru the Hong Kim model gives the correct zero and high frequency limits for the LFC.
The third order approximation involves the sixth moment of the dynamical structure factor, $<\omega^6>$; since this is difficult to compute theoretically, in the cases where the third order approximation was considered by Hong and Kim, $<\omega^6>$ was treated as an adjustable parameter \cite{Hong}.
\subsection{Computing the Dynamic Local Field Correction}

In our MD simulations, we compute directly the intermediate scattering function $F(k,t)$.  The response function is then given as
\begin{equation}
\chi(k,t) = \left\{ \begin{array}{rl}
  0 &\mbox{ if $t<0$} \\
  -\frac{n}{2 k_BT}\frac{dF(k,t)}{dt} &\mbox{ if $t\geq 0$}
       \end{array} \right.
\,,
\label{chikt}
\end{equation}
which is simply the fluctuation dissipation theorem in the temporal domain (cf. Eq. (\ref{flucdiss})).
Numerically,
we obtain the response function in the frequency domain, $\chi(k,\omega)$, by taking the discrete Fourier transform of $\chi(k,t)$.  Finally, we use the definition given in Eq. (\ref{lfcdef}) the compute the LFC.

We find that the LFC is in general rather more difficult to compute with MD than the DSF; this is reflected in the less accurate and more noisy MD data we have obtained for $G(k,\omega)$.  This is despite the fact that both $S(k,\omega)$ and $G(k,\omega)$ are derived from the same MD data for the intermediate scattering function $F(k,t)$.  In particular, it is difficult to obtain accurately the precise way in which the imaginary part of $G(k,\omega)$ decays to zero and the real part decays to its high frequency limit $G(k,\infty)$.  This is because of the way in which $G(k,\omega)$ is defined (Eq. \ref{lfcdef}): as $\omega$ increases, both $\chi_0(k,\omega)$ and $\chi(k,\omega)$ are small quantities.  Despite these difficulties, our data is sufficiently good to allow for a comparison with the models of Tanaka and Ichimaru and Hong and Kim outlined above.

\subsection{MD results and comparison to models}

 \begin{figure}
 \includegraphics{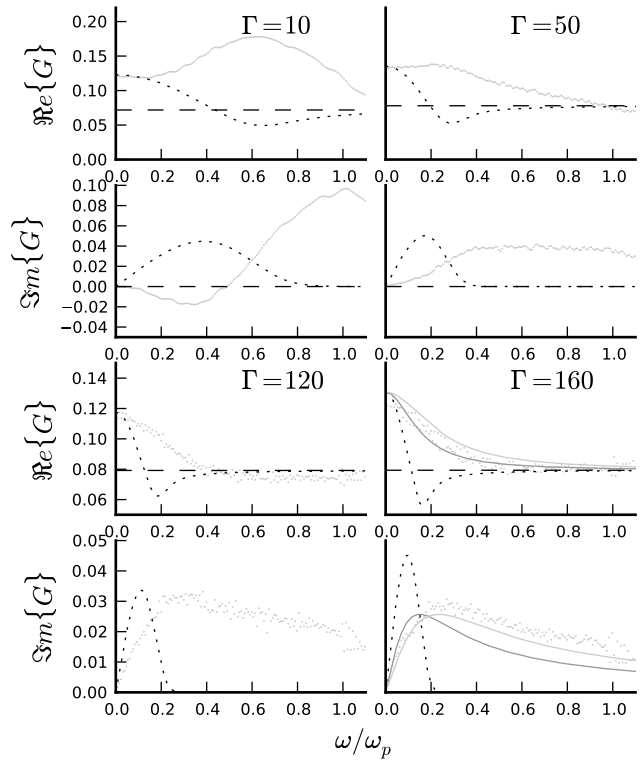}
 \caption{Real and imaginary parts of dynamic local field correction $G(k,\omega)$ as
   computed from MD at $ka = 1.02$ (dots). Also shown is the model of Hong and Kim as given in Eq. (\ref{hongkim}) (dotted line).  For $\Gamma = 160$, the model of Tanaka and Ichimaru is shown with both the Gaussian approximation (light grey line) and the Lorentzian approximation (dark grey line) for the relaxation time $\tau_M(k)$.  The dashed lines ($\Re e\{G(k,\omega)\}$ only) show $G(k,\infty)$.}
 \label{gkwsmall}
 \end{figure}

 \begin{figure}[t]
 \includegraphics{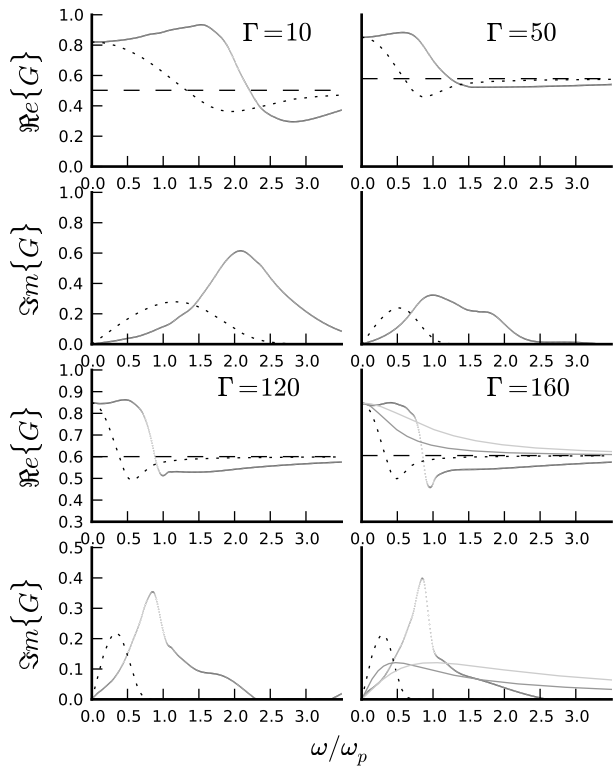}
 \caption{Same as Fig. \ref{gkwsmall} but for $ka = 3.09$.}
 \label{gkwmid}
 \end{figure}

 \begin{figure}[t]
 \includegraphics{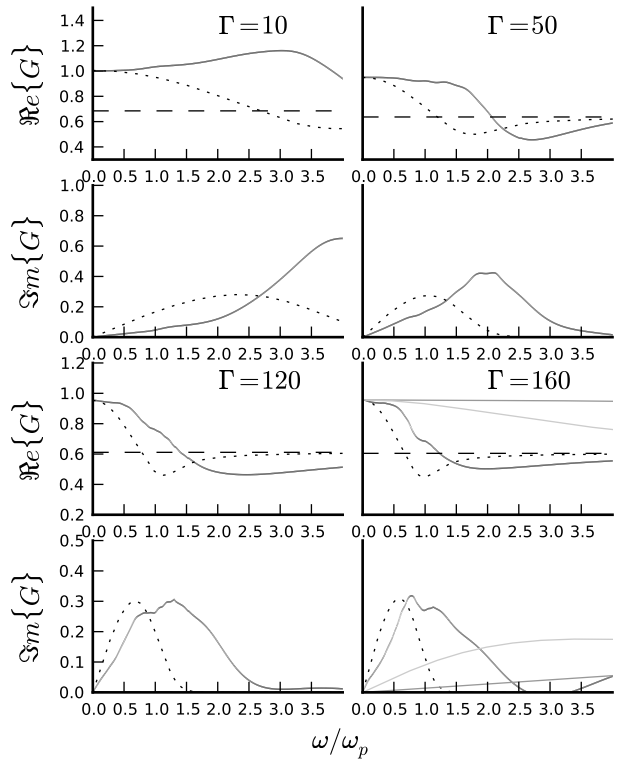}
 \caption{Same as Fig. \ref{gkwsmall} but for $ka = 6.19$.}
 \label{gkwlarge}
 \end{figure}

In Figs. \ref{gkwsmall}, \ref{gkwmid} and \ref{gkwlarge}, we have shown our MD results for the LFC at $\Gamma = 10,50,120 \mbox{ and } 160$ for small, intermediate and large $k$ respectively ($ka = 1.02, 3.09 \mbox{ and } 6.19$).
In each of these figures we also show the model of Hong and Kim (Eq. (\ref{hongkim})).  Since the relaxation time $\tau_M(k)$ appearing in the model of Tanaka and Ichimaru (Eq. (\ref{ichimarueq})) was only given explicitly for $\Gamma = 160$ (see \cite{Ichimaru}), we have shown their model - with both the Gaussian and the Lorentzian approximation for the relaxation time - for this coupling strength only
\footnote{In \cite{Ichimaru}, the Gaussian approximation is referred to as ``scheme I'', and the Lorentzian as ``scheme II''.} .

At our small $k$ value ($ka = 1.02$ - Fig. \ref{gkwsmall}), for all coupling strengths the model of Hong and Kim departs from $G(k)$ and decays to its high frequency limit $G(k,\infty)$ much faster than the MD data.  
At these long wavelengths (small $k$), the OCP dynamics occur close to $\omega = \omega_p$ ; the local field correction describes the ``collisional broadening'' of the plasmon peak neglected in the mean field approximation (c.f. Figs. \ref{fig1} and \ref{fig2})
.  Therefore, the value of $G(k,\omega)$ should only be important for $\omega$ close to $\omega_p$.
At our highest coupling strength of $\Gamma = 160$, the model of Tanaka and Ichimaru works reasonably well for either a Gaussian or Lorentzian relaxation time.  A reasonable estimate of the width of the plasmon peak is obtained, as noted previously \cite{Ichimaru}.

At our intermediate $k$ value ($ka = 3.09$ - Fig. \ref{gkwmid}), $G(k,\omega)$ shows rather more structure than at $ka = 1.02$, particularly at our largest coupling strengths of $\Gamma = 120$ and $160$.  At these coupling strengths, the sharp variation of both the real and imaginary parts of $G(k,\omega)$ around $\omega = \omega_p$ accounts for the high frequency peak in the dynamical structure factor $S(k,\omega)$ discussed in Sec \ref{skwcomp}. 
Again, for all $\Gamma$, the model of Hong and Kim departs from $G(k)$ and decays to its high frequency limit $G(k,\infty)$ much faster than does the MD data.  Furthermore, at $\Gamma = 160$, the model of Tanaka and Ichimaru cannot capture the considerable structure in $G(k,\omega)$.

At our large $k$ value ($ka = 6.19$ - Fig. \ref{gkwlarge}), again none of the models seem to give a good description of $G(k,\omega)$.

\section{Onset of negative dispersion}
\label{negdispersion}
Very recently, there has been renewed interest in the value of $\Gamma$ at
which `negative dispersion' of the OCP plasmon mode sets in  \cite{Arkhipov}. Negative
dispersion, refering to $d\omega(k)/dk < 0$, where $\omega$ is the
frequency and $k$ the wavenumber of the plasmon mode, is a feature not predicted by mean-field theory
(i.e. the Vlasov equation for the single particle distribution
function).  This anomalous effect, that represents an effect of the
`strong coupling' ($\Gamma > 1$) on the collective dynamics of the
OCP, was first discovered in early computer simulations of
the OCP \cite{Hansen}.

\subsection{MD results for plasmon peak position}
\begin{figure}[h]
\includegraphics[width=\columnwidth]{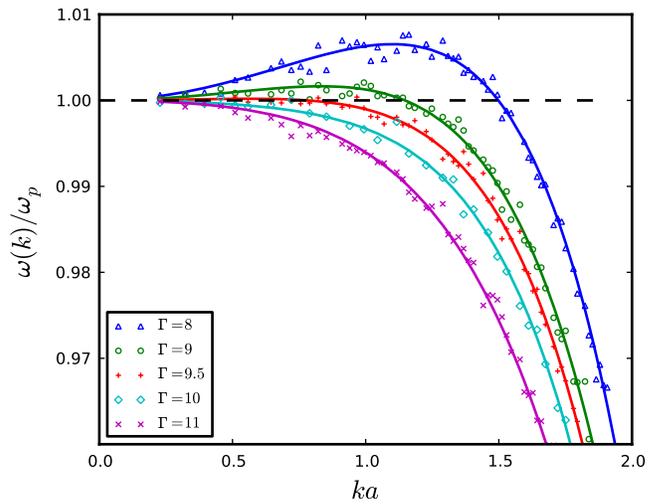}
\caption{(color online) Position of plasmon peak
  as obtained from MD simulations (open symbols) along with the least squares
fits to the functional form in Eq. (\ref{fiteq2}) with the parameters shown in Table \ref{fittable} (solid lines).}
\label{neg1}
\end{figure}

\begin{table}[t]
\begin{ruledtabular}
\begin{tabular}{lrrrrr}
& $\Gamma=8$ & $\Gamma=9$ & $\Gamma=9.5$ & $\Gamma=10$ & $\Gamma=11$ \\\hline
$b$ &0.01948 &0.00887 &0.00221 &-0.00304 &-0.00523\\ \hline
$c$ &-0.06467 &-0.06113 &-0.03124 &-0.02160 &-0.07313\\ \hline
$d$ &-0.52751 &-0.44733 &-0.59515 &-0.61923 &-0.27276\\ \hline
\end{tabular}
\end{ruledtabular}
\caption{Parameters obtained from the fit for the plasmon peak position given in Eq. (\ref{fiteq2})}
\label{fittable}
\end{table}

\begin{figure}[h]
\includegraphics[width=\columnwidth]{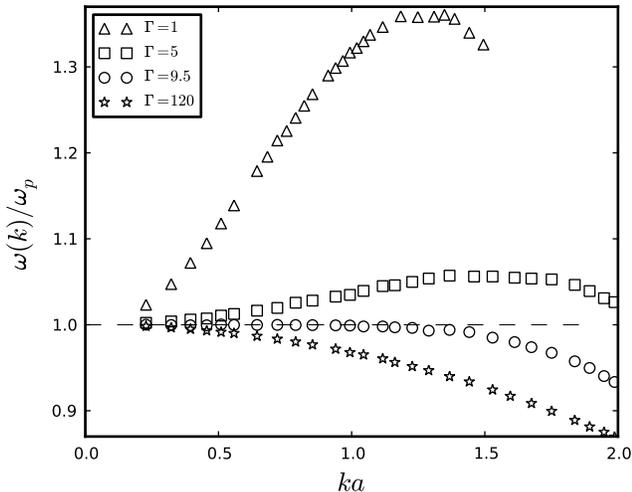}
\caption{Position of plasmon peak
  as obtained from MD simulations.}
\label{neg2}
\end{figure}

\begin{figure*}[t]
 \includegraphics{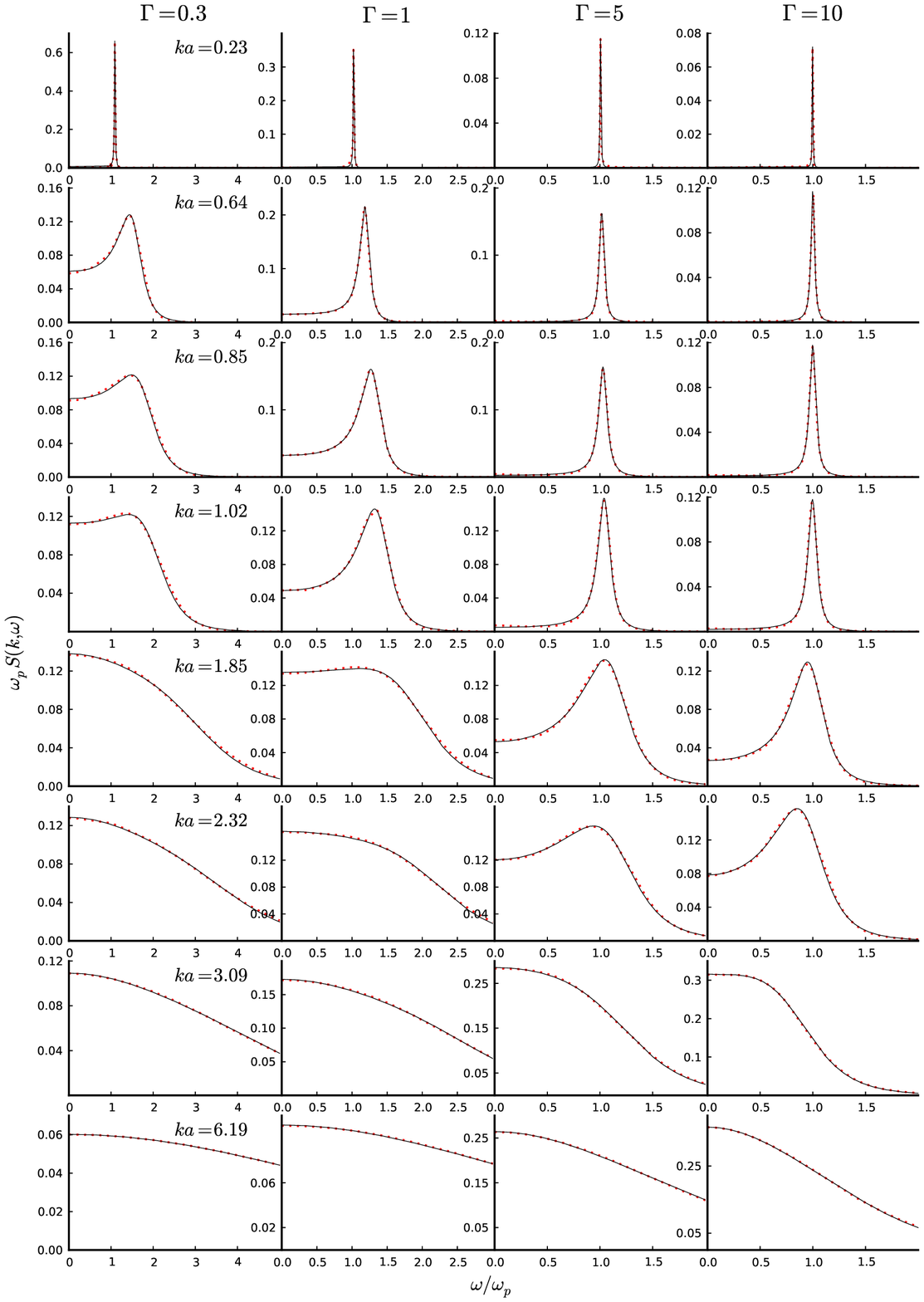}
 \caption{(color online) MD results for $S(k,\omega)$ at $\Gamma =
   0.3, 1,5$ and $10$ (dots) and the memory function fits
   (solid lines).
 }
 \label{fig1}
 \end{figure*}

\begin{figure*}[t]
 \includegraphics{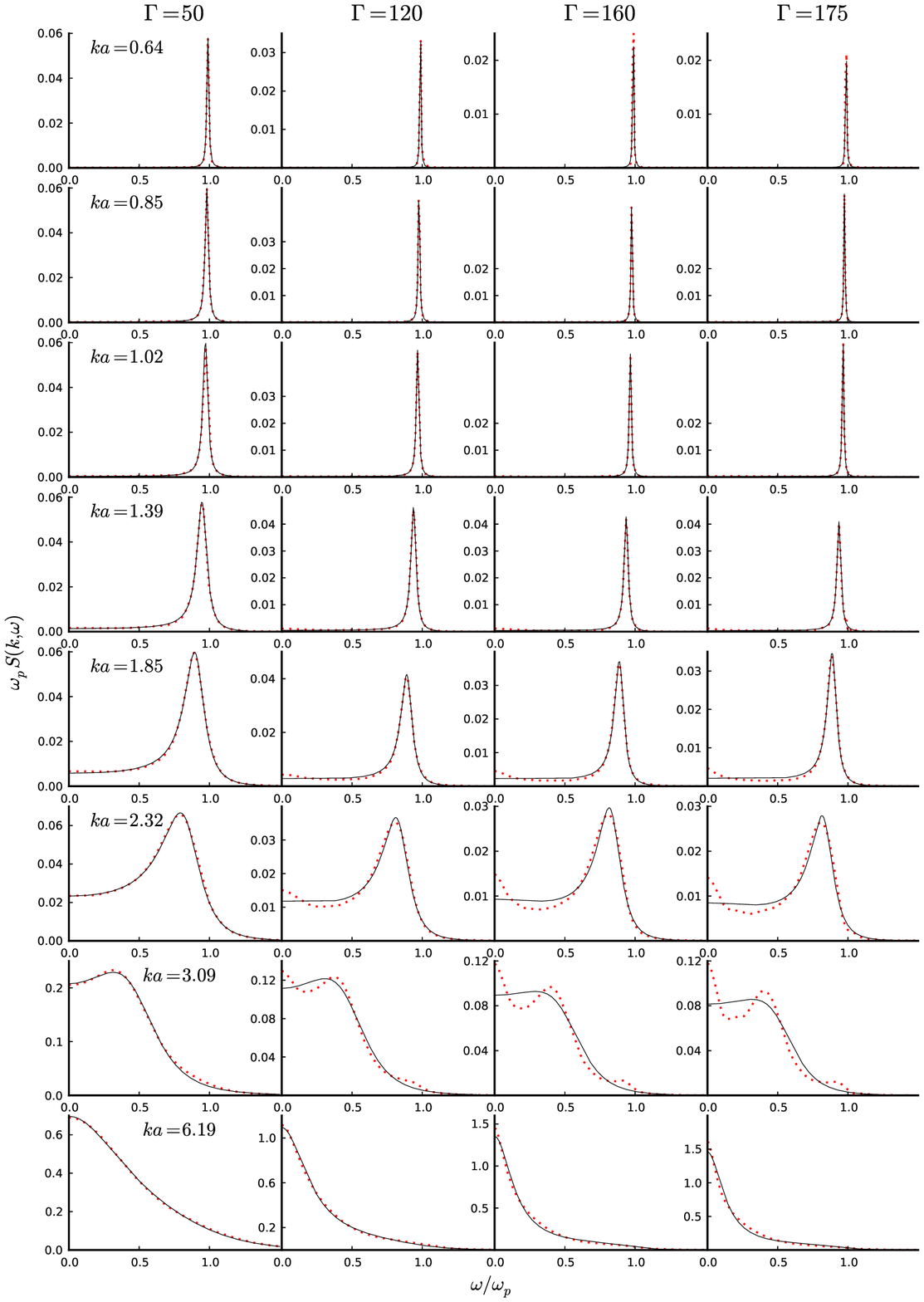}
 \caption{(color online) MD results for $S(k,\omega)$ at $\Gamma =
   50, 120,160$ and $175$ (dots) and the memory function fits
   (solid lines).
 }
 \label{fig2}
 \end{figure*}

Figure \ref{neg1} shows the plasmon peak position determined from the MD
simulations against $ka$ for coupling strengths
$\Gamma = 8,9,9.5,10,11$. 
As illustrated in Fig. \ref{neg1}, we obtained a good (least squares) fit of the MD results for $w(k)/\omega_p$ to the polynomial
\begin{equation}
\omega(k)/\omega_p = 1 + \frac{b}{2!}(ka)^2 + \frac{c}{4!}(ka)^4 + \frac{d}{6!}(ka)^6\,,
\label{fiteq2}
\end{equation}
where higher order terms in $ka$ were found to contribute negligibly to the
quality of the fit.
The values obtained for $b$,$c$, and $d$ at each coupling strength are given in
Table {\ref{fittable}}.  From the $b$ coefficient, we deduce that
negative dispersion at long wavelengths sets in between
$\Gamma = 9.5$ and $\Gamma = 10$.  Finally, in Fig. \ref{neg2}, we show the position of the plasmon peak for a larger range of $\Gamma$ values.

\section{Concluding Comments}
In this paper, we investigated two different approaches to describing the near equilibrium dynamical properties of the classical one-component plasma (OCP): the memory function, which is a standard approach for normal liquids, and the dynamic local field correction (LFC), which is more familiar to plasma physics.
Our study was centered around our highly accurate, state of the art Molecular Dynamics (MD) simulations for the intermediate scattering function $F(k,t)$.
The accuracy of our MD data allowed us to compute not only the dynamical structure factor (DSF), $S(k,\omega)$, but also the dynamic local field correction (LFC), $G(k,\omega)$, the latter of which has to our knowledge never been computed before.
We found that the memory function is rather more simple to model than the LFC: while the memory function is very well reproduced by a Gaussian for all coupling strengths $\Gamma$ and wavevectors $k$, the LFC has considerably more structure.
The more complex structure of the LFC is reflected in the fact that current models - those of Tanaka and Ichimaru \cite{Ichimaru} and Hong and Kim \cite{Hong} - do not offer a good description for a wide range of conditions.
As well as examining these two approaches, we used our MD data to accurately determine the coupling strength $\Gamma$ at which the transition from positive to negative dispersion of the plasmon mode at long wavelengths takes place, as requested by a recently published study \cite{Arkhipov}.
Aside from elucidating certain features of OCP dynamics, our MD results should find future application among practitioners in the field of strongly coupled Coulomb systems.

\appendix*

\section{Frequency moments of $S(k,\omega)$}
\label{appendix2}
The wavevector dependent quantities,
\begin{equation}
<\omega_k^2> = \frac{<\omega^2>}{<\omega^0>}\,,
\end{equation}
and
\begin{equation}
\omega_L^2(k) = \frac{<\omega^4>}{<\omega^2>}\,,
\end{equation}
are given in terms of
the frequency moments of $S(k,\omega)$, defined as
\begin{equation}
<\omega^n> = \int_{-\infty}^{\infty}\omega^nS(k,\omega)d\omega\,.
\end{equation}
The zeroth moment of $S(k,\omega)$ gives the static structure factor $S(k)$
\begin{equation}
<\omega^{0}> = S(k)\,.
\end{equation}
The second moment is
\begin{equation}
\frac{<\omega^{2}>}{\omega_p^2} = \frac{q^2}{3\Gamma}\,,
\end{equation}
where $q = ka$ with $a = (3/(4\pi n))^{1/3}$ the average interparticle spacing and $\omega_p = \sqrt{4\pi (Ze)^2n/m}$ is the plasma frequency.  
The fourth moment is \cite{Hansen}
\begin{equation}
\frac{<\omega^4>}{\omega_p^4} = \frac{1}{3\Gamma}\left[\frac{q^4}{\Gamma} + q^2 - 2q^2I(q)\right]\,,
\label{4thmoment}
\end{equation}
with
\footnote{Note that we define $I(q)$ as in \cite{Hansen}.  This quantity has a numerical value of exactly half that defined in e.g. \cite{Ichimarurev}, Eq. (3.37).}
\begin{equation}
I(q) = \int_0^{\infty}\frac{1}{\bar{r}}[g(\bar{r}) - 1]\times\left(\frac{\sin q\bar{r}}{q\bar{r}} + \frac{3\cos q\bar{r}}{(q\bar{r})^2} - \frac{3\sin q\bar{r}}{(q\bar{r})^3}\right)d\bar{r}\,,
\end{equation}
where $\bar{r} = r/a$.

\end{document}